\documentclass[aps,preprint,prd,floatfix,nofootinbib]{revtex4-1}
\pdfoutput=1
\usepackage{amsmath} 
\usepackage{graphicx}
\usepackage{color}
\newcommand{\bea}{\begin{eqnarray}} 
\newcommand{\eea}{\end{eqnarray}}

\newcommand{\WKB}{\mbox{\tiny $\sf WKB$ \normalsize}}
\newcommand{\nocontentsline}[3]{}
\newcommand{\tocless}[2]{\bgroup\let\addcontentsline=\nocontentsline#1{#2}\egroup}
\begin{document}
\title{Diving inside a hairy black hole}
\author{Nicol\'{a}s Grandi}
\email{grandi@fisica.unlp.edu.ar}
\affiliation{Instituto de F\'{i}sica de La Plata - CONICET  \&  Departamento de F\'{\i}sica - UNLP,\\ C.C. 67, 1900 La Plata, Argentina}
\author{Ignacio Salazar Landea}
\email{peznacho@gmail.com}
\affiliation{Instituto de F\'{i}sica de La Plata - CONICET  \&  Departamento de F\'{\i}sica - UNLP,\\ C.C. 67, 1900 La Plata, Argentina}
\begin{abstract}
We investigate the interior of the Einstein-Gauss-Bonnet charged black-hole with scalar hair. 
 We find a variety of dynamical epochs, with the particular important feature that the Cauchy horizon is not present. 
This makes the violation of the no-hair theorem 
a possible tool to understand how might the strong cosmic censorship conjecture  work.
\end{abstract}
\maketitle 
\tableofcontents
\newpage
\section{Introduction}
\label{sec:Introduction}
Black hole interiors are interesting and challenging from the theoretical perspective. Take for instance the charged spherically symmetric Reisser-Nordstr\"om black hole. After crossing the event horizon, a Cauchy horizon shows up before reaching the singularity. Cauchy horizons are bad, since they ruin the predictability of the classical dynamics, even if they may be away from any highly curved region \cite{Geroch:1970uw}. To deal with this, Penrose's strong cosmic censorship
conjecture posits that such Cauchy horizons are artifacts of the highly symmetric vacuum solutions
that are known analytically, and do not arise from generic initial data \cite{Simpson:1973ua}.

The evolution from arbitrary initial data is hard to actually compute, making it difficult to test the strong cosmic censorship
conjecture. To that end, an alternative approach is instead to study spherically symmetric examples perturbed with some extra matter.   

In asymptotically AdS space, a first step in such direction was done in \cite{Hartnoll:2020rwq} where a planar AdS-Reisner-Nordstrom geometry was studied. It has a Cauchy horizon, which disappears when a relevant deformation is added to the theory.
A similar situation happens when the black hole develops a scalar hair becoming the geometry known as ``holographic superconductor''  \cite{Hartnoll:2008vx,Gubser:2008px}.  Once again there is no Cauchy horizon  \cite{Hartnoll:2020fhc,Cai:2020wrp} making contact between two long standing open fields in general relativity: the aforementioned strong cosmic censorship conjecture, and the many versions of no hair theorems that started with \cite{Ruffini:1971bza}. Interestingly, the onset of the instability giving rise to the scalar hair can be generically studied in the probe limit, as it can be predicted in the linear regime where the scalar decouples \cite{Amado:2009ts,Amado:2013xya}. This implies that even when the  backreaction of the scalar hair is negligible outside the horizon, inside it something very dramatic might be happening, as the Cauchy horizon disappears. In \cite{Hartnoll:2020fhc}, interesting dynamical epochs were found and described in certain detail. 

No hair theorems make it hard to extend the above results to asymptotically flat space \cite{Ruffini:1971bza,chase,Bekenstein:1971hc,Herdeiro:2015waa}. Luckily, there are ways to bypass those theorems. A first possibility is to study non-linear matter
dynamics where the resistance of the matter field against collapse into the black hole is anchored
to these non-linearities. Such is the case of black holes with Yang-Mills \cite{Volkov:1990sva} or axionic \cite{weinberg} hair. In the later case, it was also shown that the Reisner-Nordstrom Cauchy horizon is unstable and disappears for hairy solutions. Considering the above discussion, it would be nice if the existence of hair could be predicted at the linear level.  
Such is the case for solutions for which the matter field does not inherits the spacetime
symmetries. A first example of this phenomena are Kerr black holes with scalar hair \cite{Hod:2012px,Herdeiro:2014goa}, where the metric
remains stationary and axially symmetric,  
but the matter field
is time dependent. 

Something similar could be expected for charged spherically symmetric black holes, since they usually are the poor man's toy model to rotating ones. Even if this does not seem to be the case for a minimally coupled scalar in Einstein-Maxwell theory \cite{Mayo:1996mv}, there are simple enough extensions that do present hairy solutions. That is precisely the case for charged Einstein-Gauss-Bonnet black holes in $4+1$ dimensions \cite{Grandi:2017zgz,Brihaye:2018nta}, which represent a good playground to test these ideas. This is what this paper is about.
Interestingly, these families of hairy black holes appear as the end point of a linear instability \cite{Hod:2014baa}. In that sense, solutions could exist which are arbitrarily close to the Reissner-Nordtr\"om solutions outside the black hole, but that might lose their Cauchy horizons inside\footnote{
We should remark now, that a charged black hole interior in presence of a scalar field with generic boundary conditions was already studied in \cite{Dafermos:2003vim}. Here we present a stable asymptotically flat solution that works as a toy model of the relation of the presence of hair with the strong cosmic censorship conjecture.}. 
 
This paper organizes as follows. In Section \ref{sec:EinsteinGaussBonnetTheory} we present the Einstein-Gauss-Bonnet theory in the presence of a Maxwell field and a charged scalar, and review its hairy solutions presented in \cite{Grandi:2017zgz}. In Section \ref{sec:inside} we dive inside these hairy black holes and give a numerical preview of the different dynamical epochs that appear as we approach the central singularity. In Section \ref{sec:main} we study in detail each of these epochs, giving analytic approximations to the solution for the different fields. We close in Section \ref{sec:end} with some discussions and future directions.

\newpage
\section{Scalar hair in Einstein-Gauss-Bonnet black holes}
\label{sec:EinsteinGaussBonnetTheory}
The action for Einstein-Gauss-Bonnet theory in the presence of a Maxwell field and
a charged scalar field, reads  
\begin{equation}
S  =\int d^{5}x\sqrt
{-g}\left[ \frac{1}{2} R+\frac\alpha 2  \left(  R^{2}-4R_{\mu\nu}R^{\mu\nu}+R_{\alpha\beta
\gamma\delta}R^{\alpha\beta\gamma\delta}\right)  
-\frac{1}{4} F_{\mu\nu}F^{\mu\nu}-|D_{\mu}\Phi|^{2}-m^{2}|\Phi|^{2}\right]  \ ,
\label{eq:Action}
\end{equation}
with $D_{\mu}=\nabla_{\mu}-iqA_{\mu}$. Notice that we have re-absorbed the gravitational coupling into the electric charge of the scalar $q$ by re-scaling the electromagnetic and scalar fields. Here the Gauss-Bonnet coupling $\alpha$ has mass dimension $-2$. Together with the mass of the scalar $m$, these are the only relevant couplings of the theory.

We look for spherically symmetric stationary solutions of the above defined theory with the generic form given by the Ansatz
\bea
&&
ds^{2}=-N^2f   dt^{2}+\frac{dr^2}{f}+r^{2}d\Omega_{3}^{2}\,,
\label{eq:MetricAnsatz}
\\
&&A=h\, dt \,, 
\qquad\qquad\Phi=e^{-i \omega t} \phi  \,,  
\label{eq:ScalarAnsatz}
\eea
where $N, f, h$ and $\phi$ are functions of $r$, and $\omega$ is the frequency of the scalar field. Notice that the time dependence of the scalar field is a phase, which implies that the resulting energy momentum tensor and electric current are static and ensures the compatibility of the Ansatz. The resulting equations of motion read
\bea
&& 
N'=
\frac
{2 r^3 \left(f^2 N^2 \phi'^2+\phi^2 (\omega +q h)^2\right)}
{3 f^2 N \left(4 \alpha(1\!-\!f)+r^2\right)}\,,
\label{eq:lapse}\\~
&&
f'=-\frac{fN'}{N}+\frac
{ 
	r 
	\left(
		2 N^2 
		\left(
			3(1\!-\!f)-r^2 m^2\phi^2
		\right)
		-
		r^2 h'^2
	\right)
	}
{3  N^2 \left(4 \alpha (1\!-\!f)+r^2\right)}\,,
\label{eq:radial}\\
&&
fN\left(\frac{r^3 h'}{N}\right)'
=2   q r^3  (\omega+qh)\phi^2\,,
\label{eq:maxwell}\\
&&
f N
\left(r^3 f N \phi' 
\right)'
=
r^3 \left(m^2 f N^2-(\omega +q h)^2\right) \phi\,.
\label{eq:scalar}
\eea
We are  interested in asymptotically flat space solutions. Since normalizable scalar profiles in flat space only exist for $\omega^2<m^2$ \cite{Grandi:2017zgz}, this translates into the fact that we only find regular non-zero solutions for some values of the  parameters.
On the other hand, close to the black hole horizon at $r=r_h$ we have the expansion
\bea
f &\approx & \frac{4\pi T}{N_h}(r\!- \!r_h)+\dots\,.
\label{eq:BoundaryConditionsRadialHairy}
\\
qh &\approx & qh_h + qh'_h(r\!-\!r_h)+\dots\,,
\label{eq:BoundaryConditionsGaugeHairy}
\\
N &\approx & N_h+N'_h(r\!-\!r_h)+\dots\,.
\label{eq:BoundaryConditionsLapseHairy}
\eea
Regarding the scalar equation of motion around such radius, we get the approximated form
\bea
(4\pi T)^2(r\!-\!r_h)
\left((r\!-\!r_h)\phi' 
\right)'
\approx
\left( 
4\pi T N_hm^2(r\!-\!r_h)-(\omega\!+\!qh_h)^2\right)\phi  \,.
\label{eq:NearHorizonScalar}
\eea
We focus on solutions where the metric still remains static. Moreover, we impose that the frequency corresponds to the ``superradiant" value $\omega=-qh_h$, thus the last term in the parenthesis on the right hand side is not present. The resulting equation is solved by
\bea
\phi\approx\phi_h +\frac{m^2N_h}{4\pi T} \phi_h (r\!-\!r_h) +\dots\,.
\label{eq:BoundaryConditionScalarHairy}
\eea

Looking now for the consistency of the equations of motion \eqref{eq:lapse}-\eqref{eq:scalar} at the horizon requires 
that $N_h'$ and $T$ are fixed in terms of the remaining 
near horizon parameters $h_h,h'_h,\,N_h,\,\phi_h$. The solution can then be obtained as a power series expansion, or numerically by shooting. 

\begin{figure}[htb]
\begin{center}
\includegraphics[width=7cm]{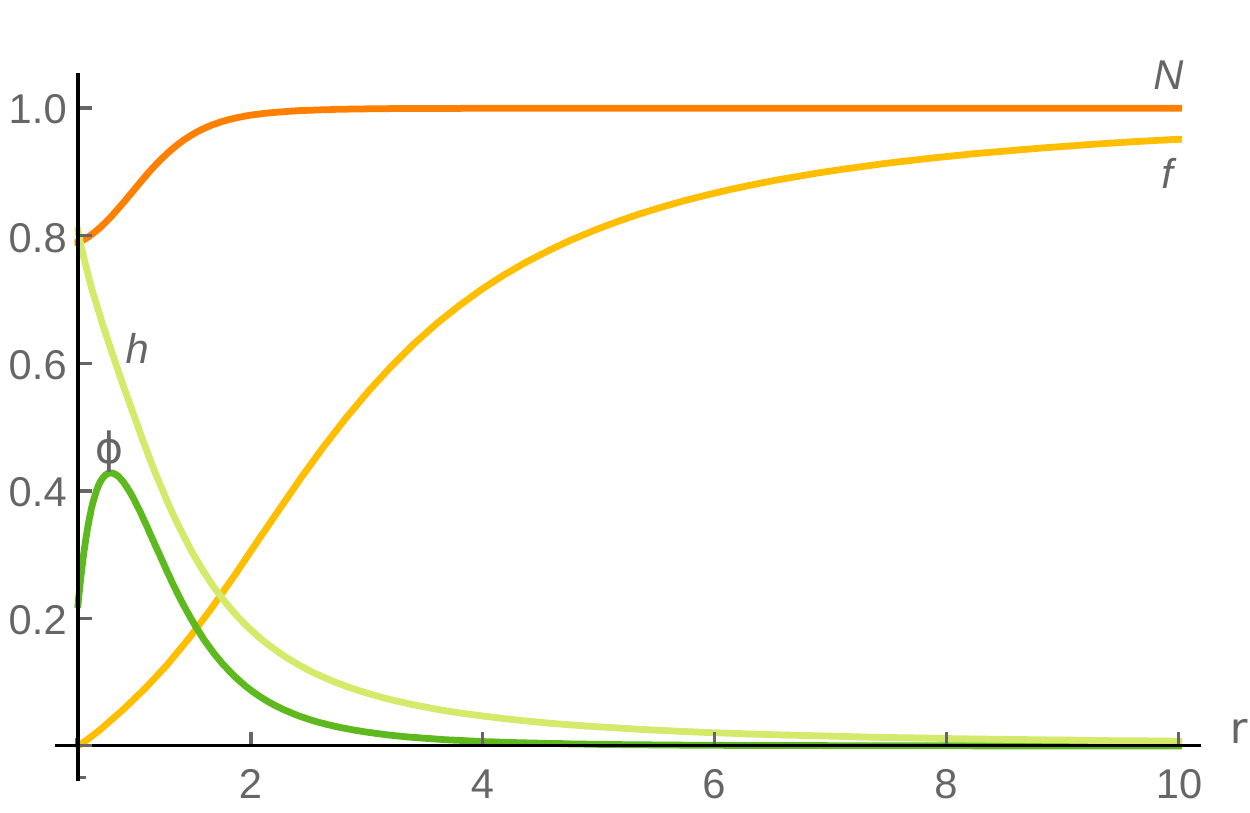}\qquad\qquad
\includegraphics[width=7cm]{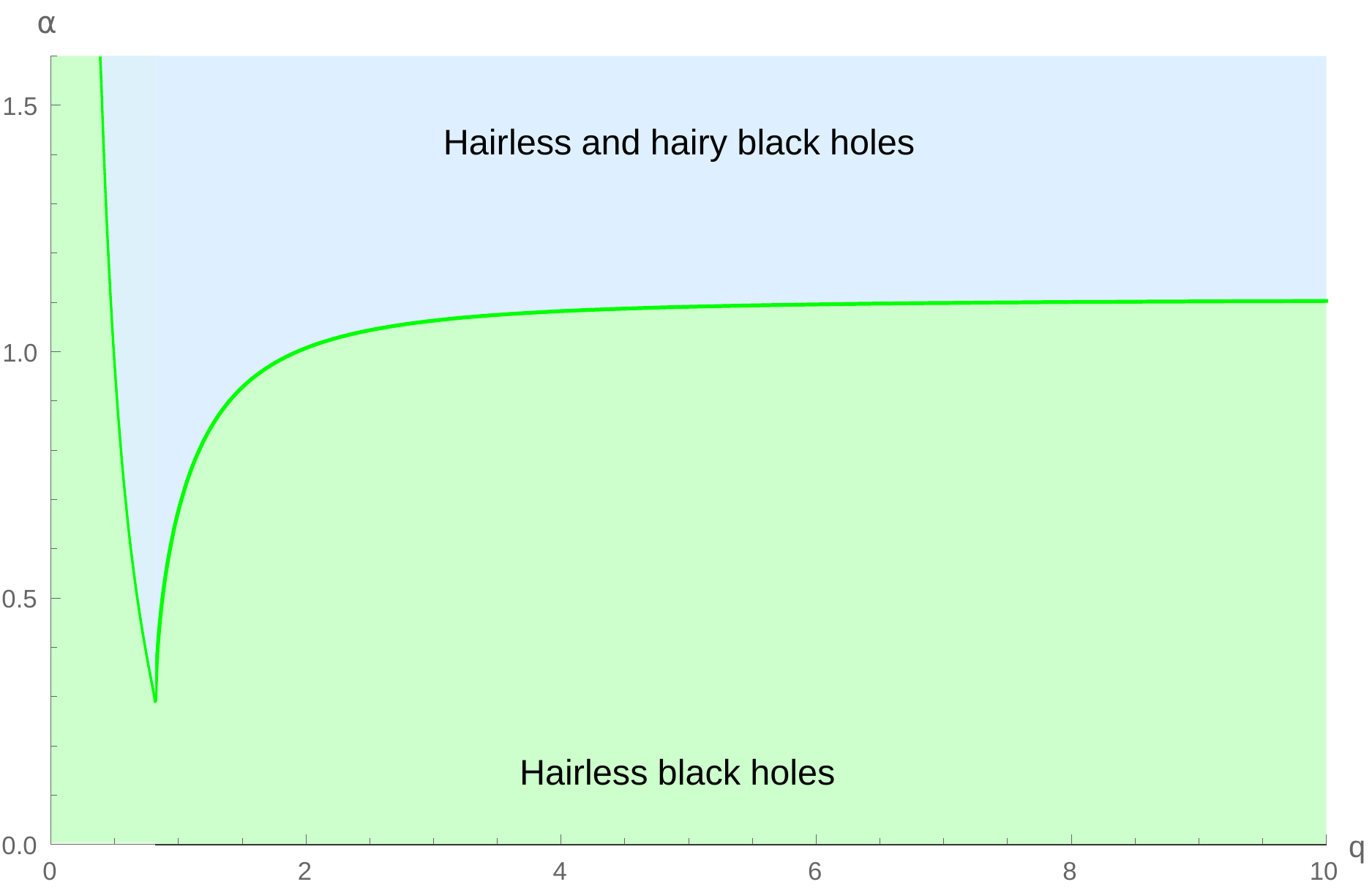}
\caption{\label{fig:HairyBlackHoleProfile}
Left: Typical profiles for the functions $N$, $f$, $h$ and $\phi$ on a hairy black hole solution, corresponding to $q=1$, $m=1$, $\alpha=2$, $h_h=0.809$, $N_h=0.790$, $h'_h=0.955$, $\phi_h=0.220$, $r_h=0.508$. Right: Regions with and without hairy black hole solutions in the $q~ vs.~\alpha$ plane.
For the sake of numerics we set $m^2=1$ and $r_h=1$.
}
\end{center}
\end{figure}

The particular case of $\phi_h=0$ reproduces the hairless Gauss-Bonnet charged black hole solution \cite{Wiltshire:1985us,Banados:2003cz}. For $\phi_h\neq 0$ hairy solutions exist \cite{Grandi:2017zgz} a typical profile of which is shown on the left of Fig. \ref{fig:HairyBlackHoleProfile}.
In the space of possible theories, spanned by the parameters $q$ and $\alpha$,  the charged hairless black hole solution exist at any point. In a particular region of parameters, scalar oscillations on such background develop a superradiant quasibound state that gives rise to the hairy black hole.  This is depicted on the right of Fig. \ref{fig:HairyBlackHoleProfile}. For a more detailed discussion, we refer the reader to \cite{Grandi:2017zgz}.

\section{Numerical preview of the hairy black hole interior}
\label{sec:inside}

Let us now fix the values of $q$ and $\alpha$ such that the hairy black hole exists. When the value of the scalar field at the horizon $\phi_h$ is small enough, we expect that the hairy solution in the exterior region  is well approximated by its probe limit, the so-called ``scalar cloud'', without affecting the metric nor the gauge field. 

\begin{figure}[h]
\begin{center}
\vspace{-.3cm}
\includegraphics[width=11cm]{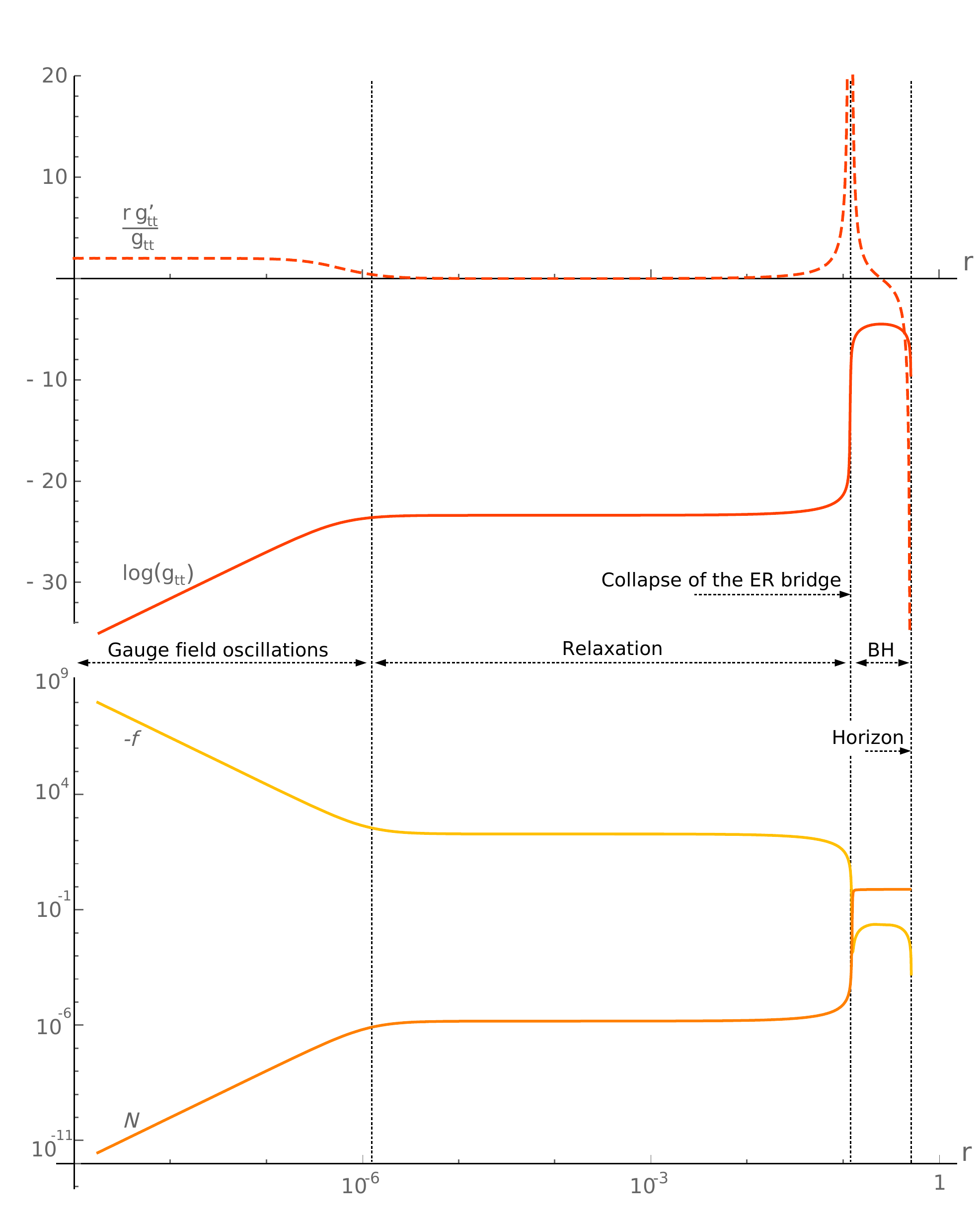}
\vspace{-.5cm}
\caption{\label{fig:prev1} Different dynamical epochs inside the horizon. Top: Profiles for $\log g_{tt}$  and ${r g_{tt}'}/{g_{tt}}$. Bottom: Profiles for $N$ and $-f$. 
This plot corresponds to $q=1,\alpha= 2,m= 1,h_h= 0.786, N_h= 0.828, h'_h=1.118, \phi_h= 0.1931, r_h=0.72$}
\end{center}
\end{figure}

Curiously, the probe approximation fails as we integrate in the opposite direction, towards the interior of the black hole. There, the effect of the scalar field becomes very important close to the position at which the Cauchy horizon would sit. Our numerics show that for an arbitrarily small amount of scalar hair, the Cauchy horizon ceases to exist and the radial coordinate remains time-like. As we keep integrating towards the singularity, a variety of epochs with qualitatively quite different behaviors emerge. To characterize them, let us look at the evolution of the metric fields as we numerically evolve the equations inside the horizon, as depicted in Figure \ref{fig:prev1}.

Close to the horizon the lapse function $N$ remains constant, while the radial function $f$ seems to follow the charged black hole solution, as it starts from zero, rises into negative values and then bounces back approaching to zero. This ``probe epoch'' lasts until we are very close to the position where we expect to find the Cauchy horizon. But there something dramatic happens, as the scalar backreaction kicks into the equations of motion, causing the lapse $N$ to drop quickly and the radial function $|f|$ to explode. This event is very sudden, and in what follows we refer to it as the ``collapse of the Einstein-Rosen bridge''.

We find useful to plot ${r g_{tt}'}/{g_{tt}}$ to characterize the power of $r$ that dominates the time component of the metric. In this way, it is easy to recognize scaling regimes where this quantity becomes approximately constant. As we see, ${r g_{tt}'}/{g_{tt}}$ blows up at the collapse of the Einstein-Rosen bridge, and immediately after that the system enters into a regime where $g_{tt}$ varies slower than any power of $r$ while the gauge fields remains constant. We call this regime the ``relaxation epoch''.

\begin{figure}[h]
\begin{center}
\vspace{-.3cm}
\includegraphics[width=8.2cm]{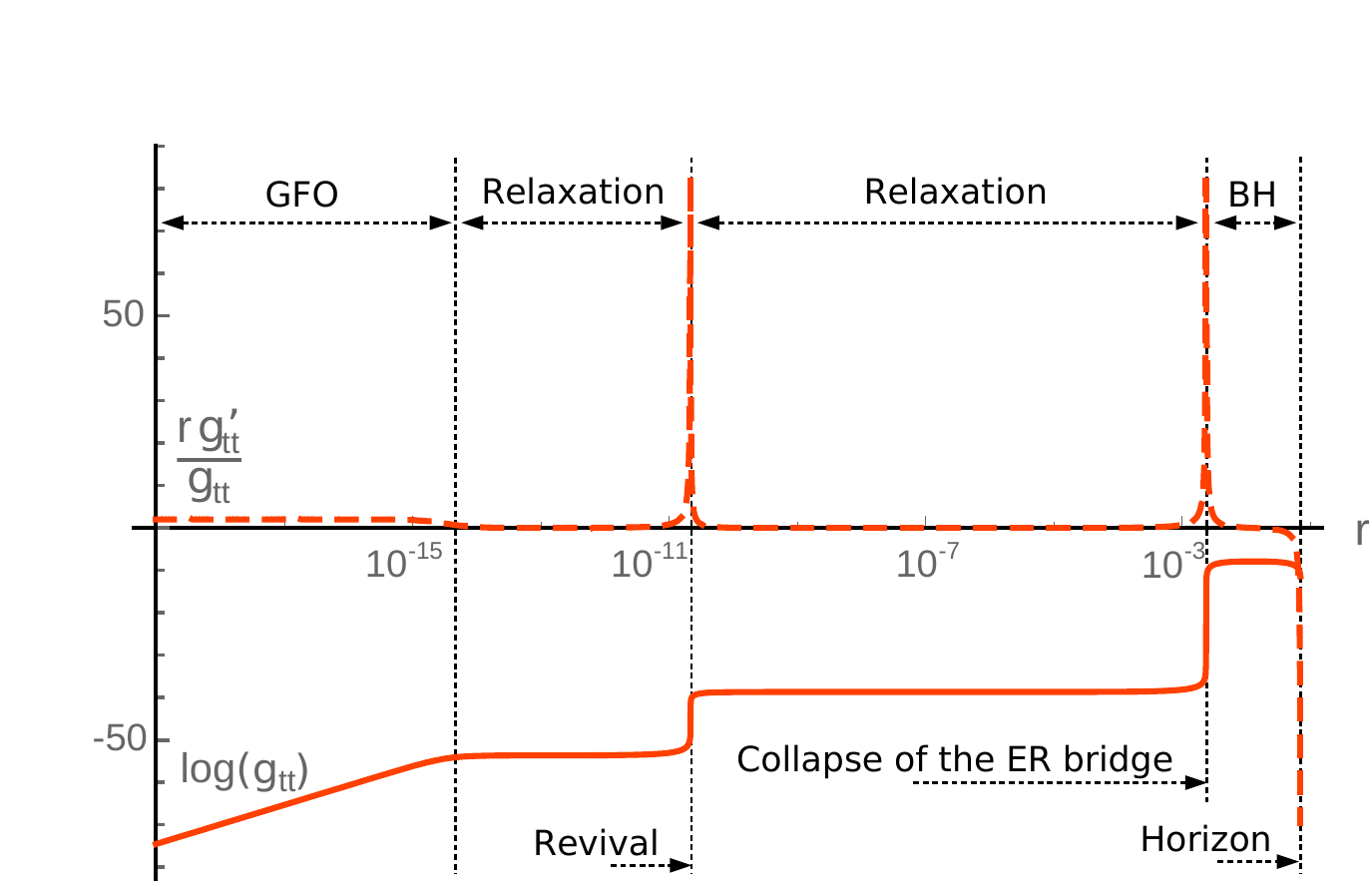}\hfill
\includegraphics[width=8.2cm]{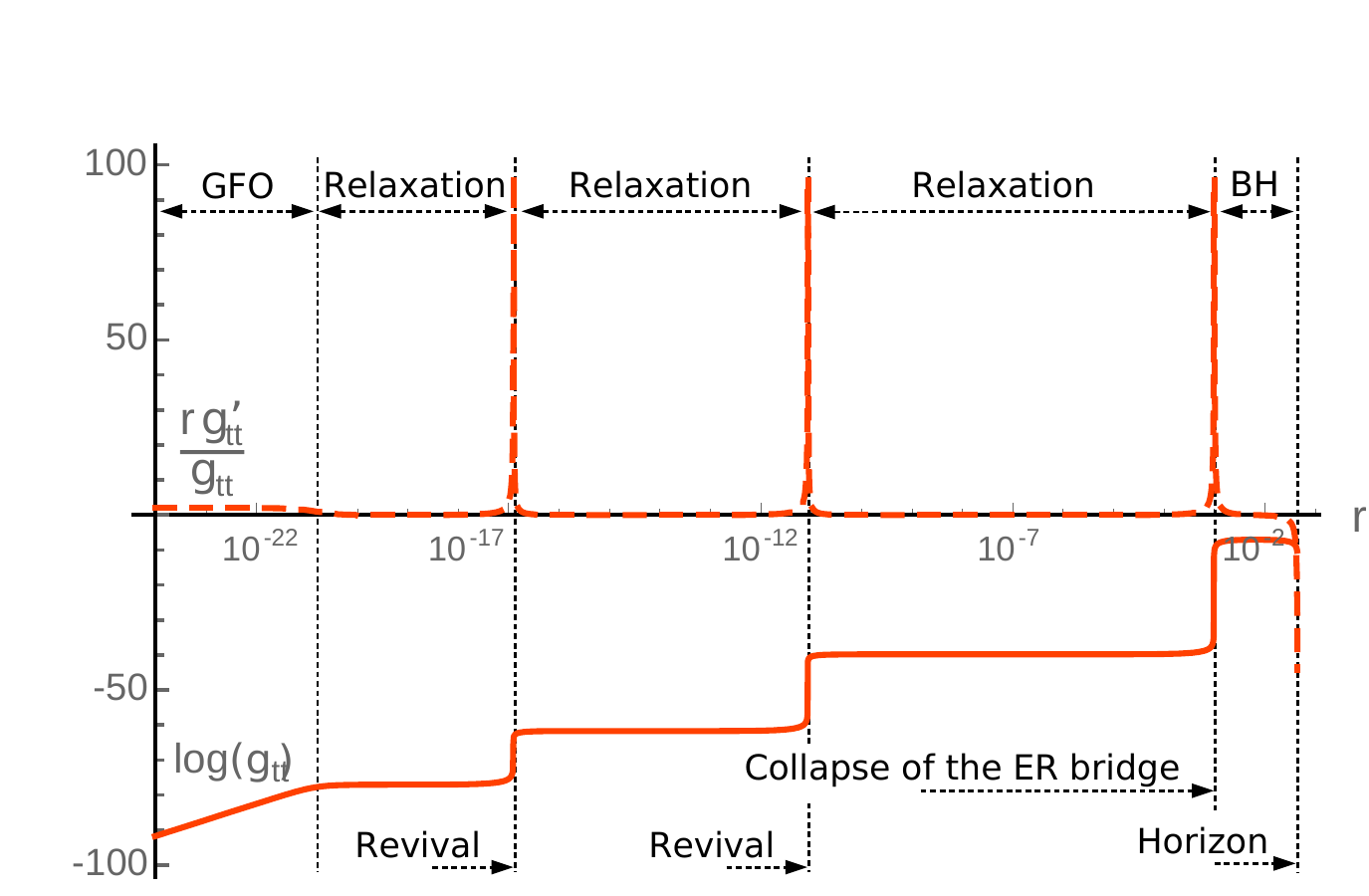}
\vspace{-.5cm}
\caption{\label{fig:revivals}    Revivals: profiles of $\log g_{tt}$  and ${r g_{tt}'}/{g_{tt}}$ for two different solutions with  $q=1,\alpha= 2,m= 1, h_h=0.833, N_h= 0.788$. The one on the left corresponds to $h'_h=
0.936, \phi_h= 0.165, r_h= 0.071$, while that on the right to
$h'_h=,0.957, \phi_h 0.173, r_h=0.046$.
}
\end{center}
\end{figure}

When the scalar field at the horizon $\phi_h$ is large enough, we find solutions with successive ``revivals'' of the probe-collapse-relaxation phenomenon.  The number of revivals increase as we increase the value of $\phi_h$. Examples of them are shown  in Figure \ref{fig:revivals}.

Finally 
 the system enters into a regime where the gauge field starts oscillating while the scalar field diverges. The imprint of this ``gauge field oscillations epoch'' is to have $g_{tt}\approx r^2$.

\section{Dynamical epochs inside the black hole}
\label{sec:main}

In this section, we look in more detail at each of the epochs sketched above. We explain the main features found on the numerical solution by performing the relevant approximations in the equations of motion.  At different values of the radial timelike variable, different approximations are valid, the validity regions sometimes overlap. We construct the approximate solution by smoothly pasting trough such regions.

We start at the beginning: as we cross the event horizon, the radial direction becomes time like and our journey towards the singularity commences.

\subsection{Early polynomial epoch}
\label{sec:et}

At the early times, we can perform a polynomial expansion of the fields around the horizon $r_h$ in powers of $r-r_h$. Using the boundary conditions given by \eqref{eq:BoundaryConditionsRadialHairy}-\eqref{eq:BoundaryConditionsGaugeHairy} and \eqref{eq:BoundaryConditionScalarHairy} in terms of the four free parameters $N_h,h_h,h'_h$ and $\phi_h$, the equations of motion  \eqref{eq:radial}-\eqref{eq:scalar} then return expressions for the higher polynomial coefficients. 

In order to paste the resulting curves to those obtained from a different approximation at a later time, we found useful to extend the polynomial expansion up to the first zero of the scalar field. For the range of $1\leq q\leq 10$ studied in the present work, the polynomial of order eight sticks pretty well to the numerical solution up to such point.

\subsection{WKB epoch}
Even if we could in principle extend the power series expansion further into the interior of the black hole by going into higher powers of $r-r_h$, 
we find more 
convenient to replace it by a different approximation in order to go beyond the first zero of the scalar field. 

To do that, we will assume that the scalar charge $q$ is large enough, so that  we can solve for the scalar field in a WKB expansion. The resulting expression for the scalar field reads
\bea
\phi=\frac{\phi_{\WKB}}{\sqrt{r^3|h-h_h|}}\cos\left(q\int dr\, \frac{|h-h_h|}{fN}+\delta_{\WKB}\right)\,,
\label{eq:SolutionsScalarWKB}
\eea
where $\phi_{\WKB}$ and $\delta_{\WKB}$ are integration constants, that are obtained from the horizon parameters $N_h,h_h,h'_h$ and $\phi_h$ by smoothly joining the full solution at some intermediate point. To obtain the expression \eqref{eq:SolutionsScalarWKB} is straitforward, after changing variables in the scalar equation  to write it in a Schr\"odinger-like form. Notice that the approximation breaks down close to the horizon where $h=h_h$ and in the central region close to $r=0$. 

In order to obtain an expression for the metric functions $N,f$ and the gauge field $h$, we plug back \eqref{eq:SolutionsScalarWKB} into the equations \eqref{eq:lapse}-\eqref{eq:maxwell}, and keep in the resulting expressions only the highest powers of $q$, to obtain
\bea
&& 
N'=
\frac
{2  q^2\phi^2_{\WKB} |h-h_h|}
{3 f^2 N \left(4 \alpha(1\!-\!f)+r^2\right)} +{\cal O}\left(q^0\right)\,,
\label{eq:lapseWKB}\\
&&
f'=-\frac{fN'}{N}+\frac
{
		r|h-h_h|
		\left(
			6 N^2 (1\!-\!f)
			-
			r^2 h'^2
		\right)
		-
		2 N^2  {m^2\phi^2_{\WKB}}  \cos^2\!\left(\cdots\right)
	}
{3 |h-h_h| N^2 \left(4 \alpha (1\!-\!f)+r^2\right)} \,,
\label{eq:radialWKB}\\
&&
fN\left(\frac{r^3 h'}{N}\right)'
=2   q^2\phi_{\WKB}^2 \cos^2\!\left(\cdots\right) \,.
\label{eq:maxwellWKB}
\eea
These are still coupled highly non-linear equations, but as we show in the forthcoming sections, their solution can be approximated in the interior regions by analytic expressions.

\subsubsection{Probe epoch}

It is natural to assume that, for very small values of the scalar field at the horizon $\phi_h$, the value of $\phi_{\WKB}$  resulting from pasting the WKB form of the scalar field to the corresponding polynomial solution will also be very small. In such case, we can drop the right hand side on equation \eqref{eq:maxwellWKB} as well as the last term in the numerator of \eqref{eq:radialWKB}. Regarding the right hand side of \eqref{eq:lapseWKB}, we can also drop it as long as we are away of any of the zeroes of the denominator. Then we get
\bea
&& 
N'=
0+\,{\cal O}\!\left(\frac 1 f\,\frac{q^2\phi_{\WKB}^2}{fN}\right)\,,
\label{eq:lapseBH}\\
&&
f'=\frac
{
		r 
		\left(
			6 N^2 (1\!-\!f)
			-
			r^2 h'^2
		\right)
	}
{3  N^2 \left(4 \alpha (1\!-\!f)+r^2\right)}+{\cal O}\!\left(m^2\phi_{\WKB}^2\right)\,,
\label{eq:radialBH}\\
&&
 \left(\frac{r^3 h'}{N}\right)'
=0+{\cal O}\!\left(\frac{q^2\phi_{\WKB}^2}{fN}\right)\,.
\label{eq:maxwellBH}
\eea
These equations correspond to the vaccum Einstein-Gauss-Bonnet-Maxwell system, and have an analytic solution in the form of a charged black hole \cite{Wiltshire:1985us} 
\begin{align}
N=&\, N_{0}\,,
\label{eq:solWKNlapse}\\
f=& 1+\frac{r^2}{4\alpha}\left(
1- \sqrt{1+\alpha\left(\frac{8M_{0}}{r^4}-\frac{16 Q_{0}^2}{N_0^2r^6}\right)}
\right)
\,,
\label{eq:solWKNradial}\\
h=& \mu_{0}-\frac{{Q_{0}}}{r^2}\,.
\label{eq:solWKNmaxwell}
\end{align}
The integration constants ${N}_{0},{M}_{0},{Q}_{0}$ and $\mu_{0}$, together with $\phi_{\WKB}$ and $\delta_{\WKB}$ are obtained in terms of the horizon parameters $N_h, h_h,h'_h,\phi_h$ by smoothly pasting to the polinomial solution at an intermediate point. We find practical to do it at the position of the first zero of the scalar field. It is worth to mention that  ${N}_{0},{M}_{0},{Q}_{0}$ and $\mu_{0}$ are just constants of integration for an approximate form of the solution inside the horizon, and do not correspond to the ADM charges associated to the hairy black hole, which are instead defined at infinity. 

\begin{figure}[htb]
\vspace{.8cm}
\begin{center}
\includegraphics[width=7.7cm, height=5.3cm]{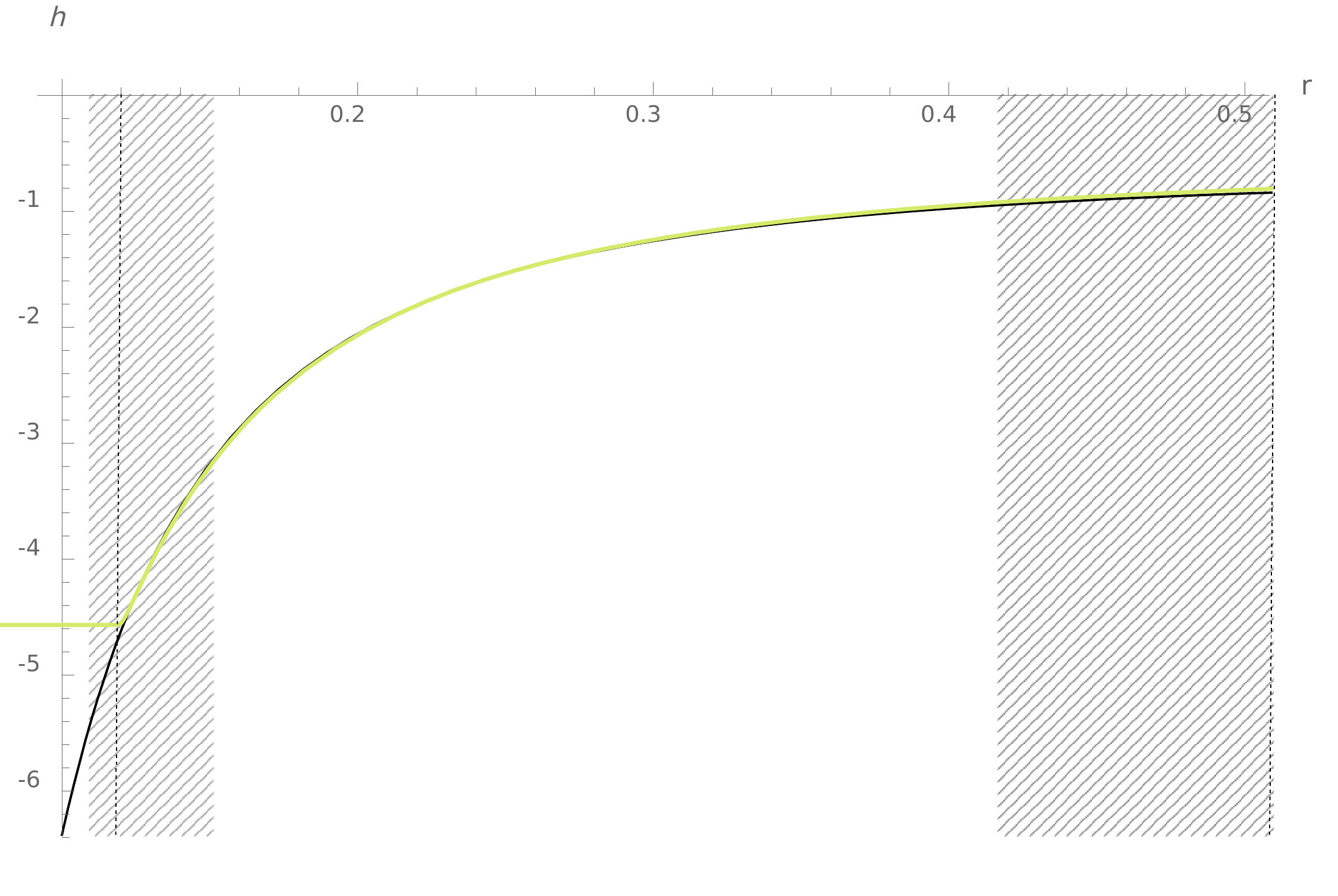}\hfill 
\includegraphics[width=7.7cm, height=5.3cm]{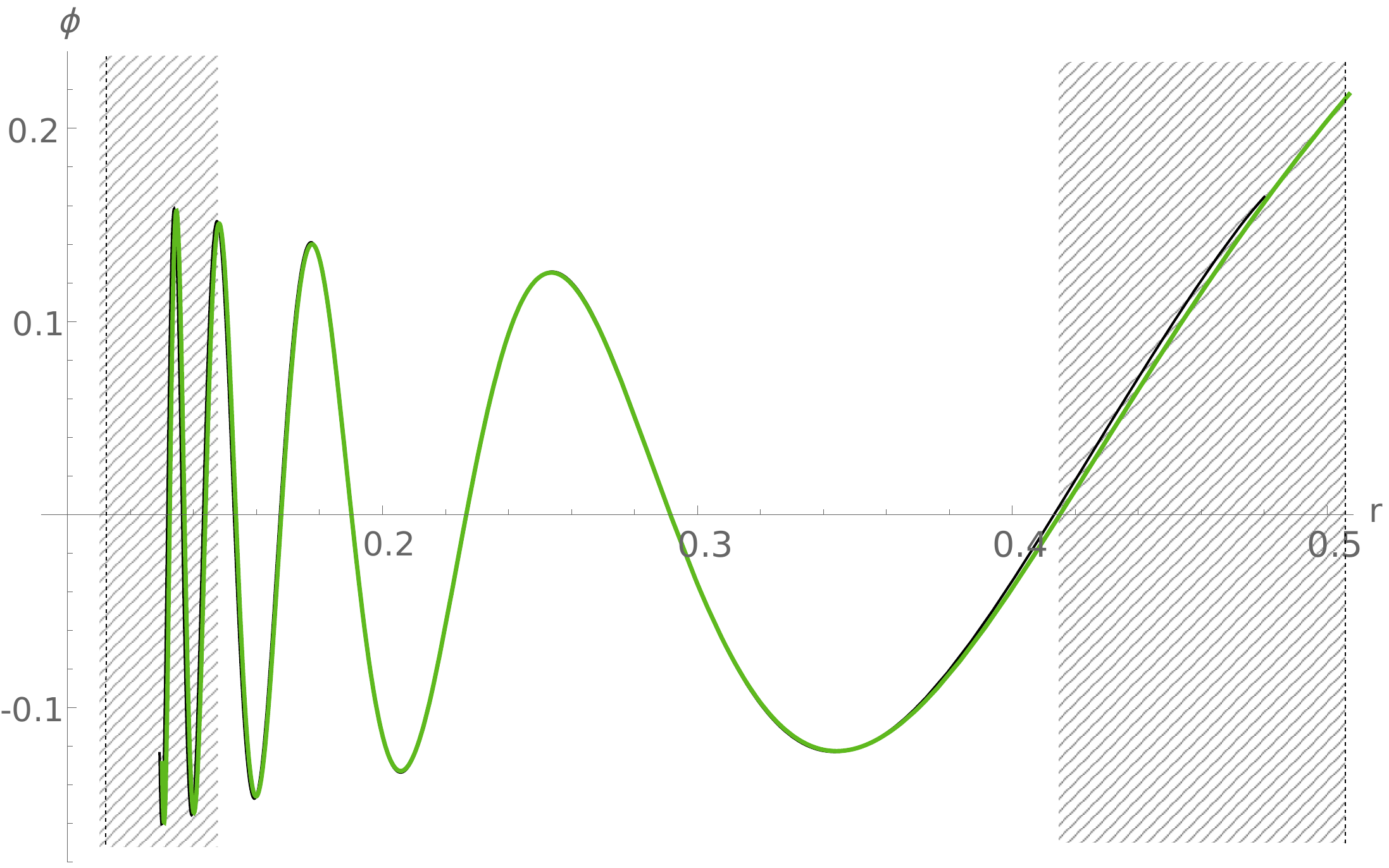}\\ \vspace{8mm}
\includegraphics[width=7.7cm, height=5.3cm]{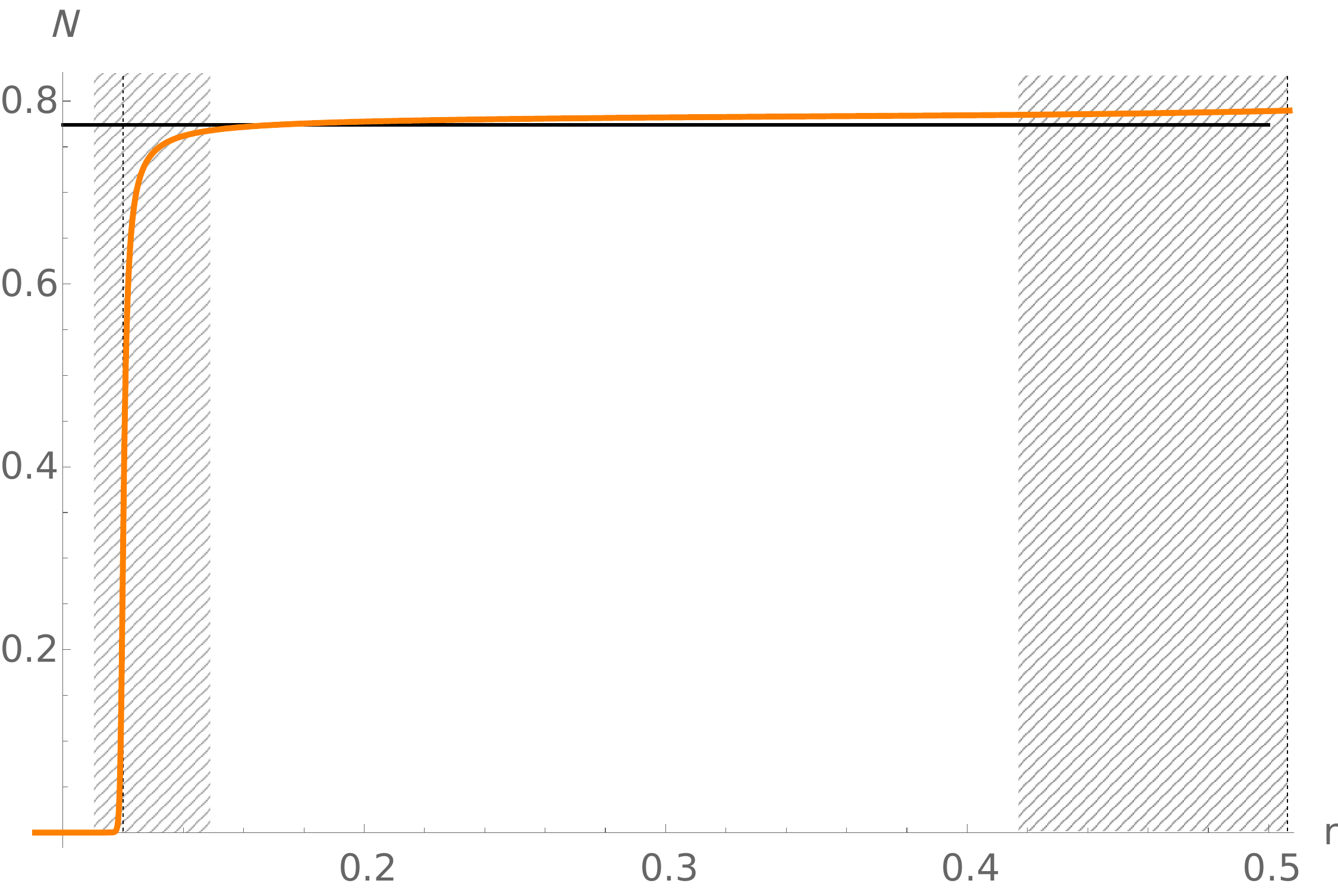}\hfill 
\includegraphics[width=8.1cm, height=5.3cm]{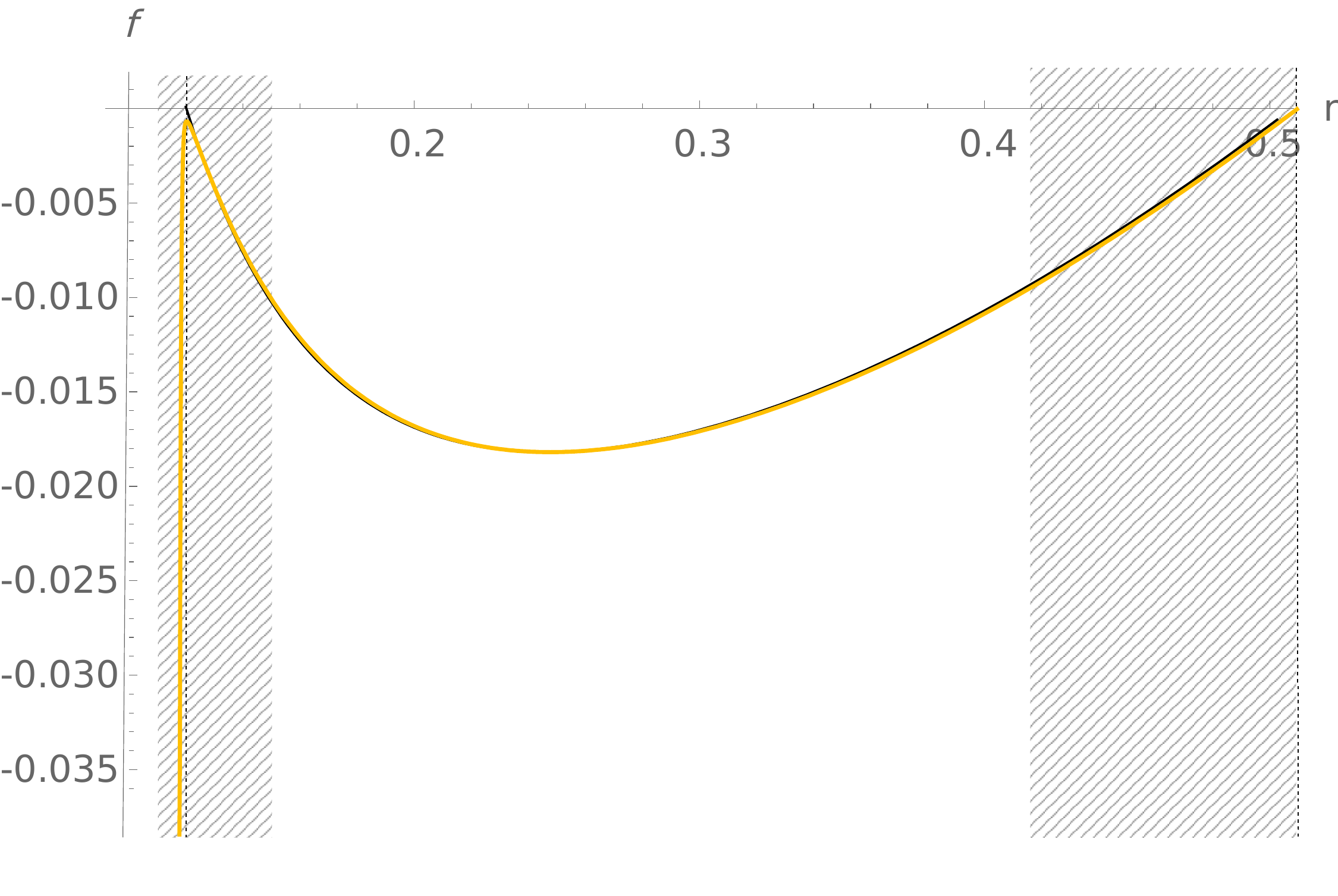}
\vspace{.5cm}
\caption{\label{fig:probe} Full non linear numerical evolution (colour) and probe approximations \eqref{eq:SolutionsScalarWKB} and \eqref{eq:solWKNlapse}-\eqref{eq:solWKNmaxwell} (black) of the profiles in the probe epoch. The dashed regions have been excluded from the fitting. That on the right of each plot interpolates to the polynomial approximation of the previous subsection, while the one on the left connects to the collapse of the Einstein-Rosen bridge described below. The dashed line at large $r$ corresponds to the black hole horizon, while that at small $r$ signals the position of the would-be Cauchy horizon. The values of the parameters correspond to the black hole shown in Fig. \ref{fig:HairyBlackHoleProfile} (left).}
\end{center}
\end{figure}

Plugging in the horizon values, we can check for how good and for how long is valid this approximation, by comparing it to the numerical plots in  Fig. \ref{fig:probe}. Notice that in our plots we have $q^2=m^2$, which lies close to the minimal value $q^2=\frac23 m^2$ needed to find hairy solutions  \cite{Grandi:2017zgz}, and  we are working with a value of $\phi_h$ relatively large. 
In this regime the dynamics outside the horizon is highly back-reacting. However, as we can see from the figures, the probe limit is still good bellow the horizon for quite a long time. 

~

An important feature of the black hole solution \eqref{eq:solWKNlapse}-\eqref{eq:solWKNmaxwell} is the presence of a Cauchy horizon at $r=r_0$ with $2r_0=2\alpha-{M_{0}}-\sqrt{\left(2\alpha-{M_{0}}\right)^2-{8 Q_{0}^2}/{3} }$, at which $f(r_0)=0$. As we approach $r_0$, the function $f$ gets smaller and the approximation we made to go from equation \eqref{eq:lapseWKB} to equation \eqref{eq:lapseBH} breaks down, due to the $1/f$ factor in the discarded terms in \eqref{eq:lapseBH}. Then the charged black hole solution \eqref{eq:solWKNlapse}-\eqref{eq:solWKNmaxwell} ceases to be a good approximation of the full solution. 

~

\newpage

~

\subsubsection{Collapse of the Einstein-Rosen bridge}
In the vicinity of its Cauchy horizon, the black-hole \eqref{eq:solWKNlapse}-\eqref{eq:solWKNmaxwell} is not a good approximation of the full solution. However, we can still drop higher powers of $\phi_{\WKB}$ in \eqref{eq:radialWKB} and \eqref{eq:maxwellWKB}. Assuming $f\ll1$ we are left with
\bea
&& 
N'=
\frac
{2  q^2\phi^2_{\WKB} |h-h_h|}
{3 f^2 N \left(4 \alpha +r^2\right)}+{\cal O}\!\left(f\right)\,,
\label{eq:lapseCollapse}\\
&&
f'=-\frac{fN'}{N}+\frac
{
		r
		\left(
			6 N^2  
			-
			r^2 h'^2
		\right)
			}
{3 N^2 \left(4 \alpha  +r^2\right)}+{\cal O}\!\left(f,m^2\phi_{\WKB}^2\right)\,,
\label{eq:radialCollapse}\\
&&
 \left(\frac{r^3 h'}{N}\right)'
=0+{\cal O}\!\left(\frac{q^2\phi_{\WKB}^2}{fN}\right)\,.
\label{eq:maxwellCollapse}
\eea
The last equation can be integrated once as $h'=(Q_{0}/2N_{0}r^3)N$, where the integration constant was chosen so as to match with the solution \eqref{eq:solWKNlapse}-\eqref{eq:solWKNmaxwell} as $r\gg r_0$. 

An important point to be noticed about \eqref{eq:lapseCollapse}-\eqref{eq:maxwellCollapse} is that this form of the equations is valid in the region around $r_0$ where $f\ll q^2\phi^2_{\WKB}/fN$. If $\phi_{\WKB}$ is small enough, as it happens for very small $\phi_h$, such region is very narrow. Writing $r=r_0+\delta r$ we can expand the remaining equations to the first order in $\delta r$, and write $f, N$  and $h$ as functions of $\delta r$. We thus obtain
\bea
&& 
f (-fN^2)_{\delta r}=(-fN^2)\frac
{
		r_0
		\left(
			6   
			-
			(Q_{0}/2N_{0}r_0^2)^2 
		\right)
			}
{3   \left(4 \alpha  +r_0^2\right)}-
\frac
{2  q^2\phi^2_{\WKB} |h -h_h|}
{3 \left(4 \alpha +r^2_0\right)}\,,
\label{eq:lapseCollapseFixed}\\
&&
f_{\delta r}=\frac
{
		r_0
		\left(
			6   
			-
			(Q_{0}/2N_{0}r_0^2)^2 
		\right)
			}
{3   \left(4 \alpha  +r_0^2\right)}
+
\frac
{2  q^2\phi^2_{\WKB} |h -h_h|}
{(-fN^2)\, 3 \left(4 \alpha +r^2_0\right)}\,,
\label{eq:radialCollapseFixed}
\eea
where we have reordered the equations in order to make explicit the temporal component of the metric $g_{tt}=-fN^2$. Taking one further derivative in the first equation, and dropping derivatives of $h$ since it behaves smoothly at $r\sim r_0$, we can obtain a single second order equation for $g_{tt}$ with the form
\bea
&& 
\frac{(g_{tt})_{\delta r\delta r}}{(g_{tt})_{\delta r}}=
\frac{ (g_{tt})_{\delta r}}{g_{tt}}
\,
\frac{1}{1+A\, g_{tt}}
\,,
\label{eq:gttCollapseFixed} 
\eea
where we defined the constant $A= \left(Q^2_{0}-24 \,r_0^4 N^2_{0} \right) /8 \,r_0^3 (q N_{0}\phi_{\WKB})^2 |h-h_h|_{r=r_0}$. This equation can be integrated into the algebraic expression
\bea
&& 
a_0^2 (r-r_0')=A \,g_{tt}+\log{g_{tt}} 
\,,
\label{eq:gttCollapseSolved} 
\eea
where $a_0^2>0$ and $r_0'$ are constants of integration. If $g_{tt}$ is large enough, we can drop the logarithm resulting in a linear behavior $g_{tt}=a_0^2(r-r_0')/A$. We can now use \eqref{eq:lapseCollapseFixed} to obtain $N$ and $f$ and match them with the charged black hole forms \eqref{eq:solWKNlapse} and  \eqref{eq:solWKNradial} by adjusting $a_0^2$ and $r_{0}'$. On the other hand, in the region where $g_{tt}$ is very small we can drop the linear term on the right and solve as $g_{tt}=\exp(a_0^2(r-r_0'))$. To be consistent, we need that $r<r_0'$, implying that we are beyond the collapse point.
\begin{figure}[tb]
\begin{center}
\includegraphics[width=10.5cm]{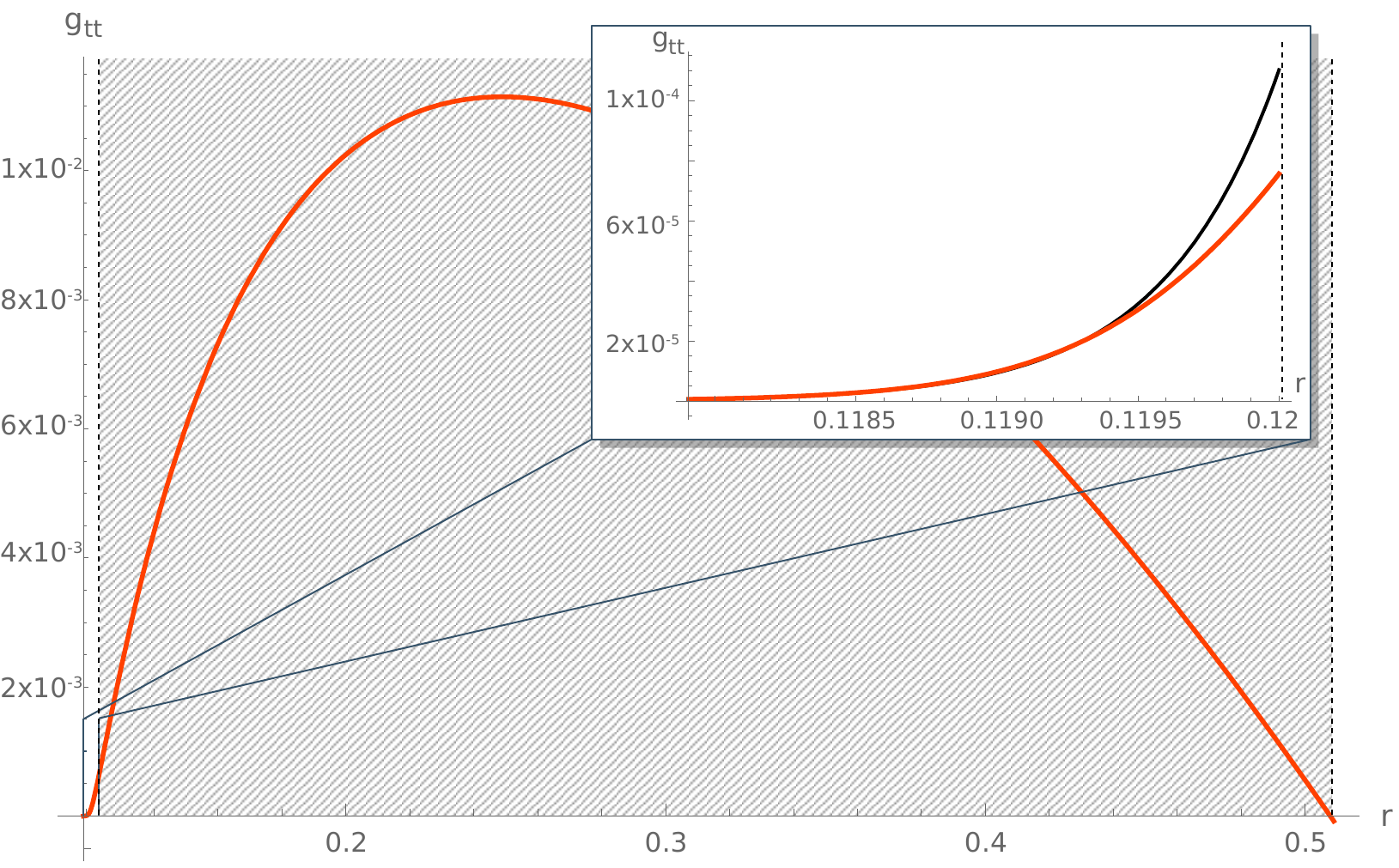}
\vspace{.3cm}
\caption{\label{fig:collapse} Full non linear evolution (color) and approximate solution (black) for $g_{tt}$ close to the collapse of the Einstein-Rose bridge. The dashed region is excluded from the fit, and corresponds to the probe and polynomial epochs mentioned in the previous subsections.
}
\end{center}
\end{figure}

In other words, as we go through the collapse point $r_0'$ we evolve from a linear behavior of $g_{tt}$ for $r>r_0'$, as if the metric where approaching a Cauchy horizon, into a non-negative but exponentially small value for $r<r_0'$. We conclude that the Cauchy horizon has dissipated.

Curiously, within these approximations the Gauss-Bonnet parameter decoupled from the relevant dynamics. It would be interesting to understand if this generically implies that the collapse of the Einstein-Rose bridge is governed by Einstein gravity dynamics, but so far its just an speculation.

Notice that $g_{tt}$ decreases exponentially as we pass the collapse point towards the singularity. Then \eqref{eq:lapseCollapseFixed} implies that $f\sim1/g_{tt}$ grows exponentially and $N\sim 1/f$ decreases exponentially.  Then we exit the collapse with a very large value of $f$ and a very small value of $N$, while the value of $fN$ is of order one.

A plot of the profile of $g_{tt}$ can be seen in Fig. \ref{fig:collapse}, in which the sudden nature of the process can be clearly observed.

\subsubsection{Relaxation}
\label{relaxation}

After the collapse, the absolute value of the radial function $f$ is very large, then in equation \eqref{eq:lapseWKB} we do not have the problem of a vanishing denominator. Nevertheless, the lapse function $N$ is now very small, so the right hand side cannot be discarded even if $\phi_{\WKB}\ll1$. With these considerations equations \eqref{eq:lapseWKB}-\eqref{eq:maxwellWKB} become
\bea
&& 
N'=-
\frac
{  q^2\phi^2_{\WKB} |h-h_h|}
{6 \alpha f^3 N }+{\cal O}\!\left(\frac{4\alpha+r^2}{4\alpha f}\right)\,,
\label{eq:lapseRel}\\
&&
(fN)'= 0+N{\cal O}\!\left(m^2\phi_{\sf WKB}^2,\frac{r^3h'^2}{N^4 },\frac{r(4\alpha+r^2)}{4\alpha f}
\right)\,,
\label{eq:radialRel}\\
&&
\left(\frac{r^3 h'}{N}\right)'
=0+{\cal O}\!\left(\frac{q^2\phi_{\WKB}^2}{fN}\right)\,.
\label{eq:maxwellRel}
\eea
The last equation could be solved as before as $h'\propto N/r^3$, but since $N$ is tiny, as long as $r$ is not too small a good approximation is $h'=0$ and then $h=h_{0}$ constant. Notice that this assumption is consistent with discarding the term with $r^2h'^2$ in going from \eqref{eq:radialWKB} to \eqref{eq:radialRel} and would be violated at smaller $r$. On the other hand, the second equation gives $fN=b_0$ in terms of a new constant $b_0$. The remaining equation can be reordered as
\bea
&& 
\left(\frac 1N\right)'=
\frac 1 B
\,,
\label{eq:lapseRel2}
\eea
where  the constant $B$ can be written as $B=
{6 \alpha b_0^3 }/{ q^2\phi^2_{\WKB} }|h_{0}-h_h|$. This can be integrated allowing us to write the complete solution %
\bea
&&
N=\frac B{r-r_{0}''}\,,
\\&&
f=\frac {b_0} B(r-r_{0}'')\,,
\\&&
h=h_{0}\,,
\eea
where $r_{0}''$ is another constant of integration.

As we keep on moving into small $r$ one of two things may happen. Either we enter into a dynamical epoch in which the scalar decouples again from the background, causing what we call a ``revival'', or we exit the region of validity of the WKB approximation, entering what we call the ``gauge field oscillations epoch''. Both possibilities are explained below.

\subsubsection{Revivals}
\label{revivals}
After the relaxation, we eventually arrive to an epoch in which the value of $r$ is very small. Provided we are still in the region of validity of the WKB approximation, the approximation of constant $h$ we made in the previous subsection to solve \eqref{eq:maxwellRel} is no longer valid. Moreover, the function $f$ has grown enough such that we can now discard the right hand side of \eqref{eq:lapseWKB}. 
In other words, we can repeat the steps taking us from \eqref{eq:lapseWKB}-\eqref{eq:maxwellWKB} to \eqref{eq:lapseBH}-\eqref{eq:maxwellBH} and to the corresponding charged black hole solution \eqref{eq:solWKNlapse}-\eqref{eq:solWKNmaxwell}, now with different values for the integration constants ${N}_1,{M}_1,{Q}_1$ and $\mu_1$.

This new black hole would have its own Cauchy horizon at $r=r_{1}$, where the approximation of large $f$ breaks down, in which case the equations to solve are instead \eqref{eq:lapseCollapse}-\eqref{eq:maxwellCollapse}. Then we are lead to a second collapse of the Einstein-Rosen bridge at a smaller radius $r_1'$. 
 As we emerge from such collapse, we are again in a relaxation region in which the solution behaves as in the previous subsection, now governed by a different radius $r''_1$ and constants $b_1$ and $h_1$, that eventually leads to a new revival. 
The process can repeat a certain number of times $n$, until $r$ is small enough and we are no longer inside the region of validity of the WKB approximation. 

~

\begin{figure}[ht]
\begin{center}
\includegraphics[width=16.3cm]{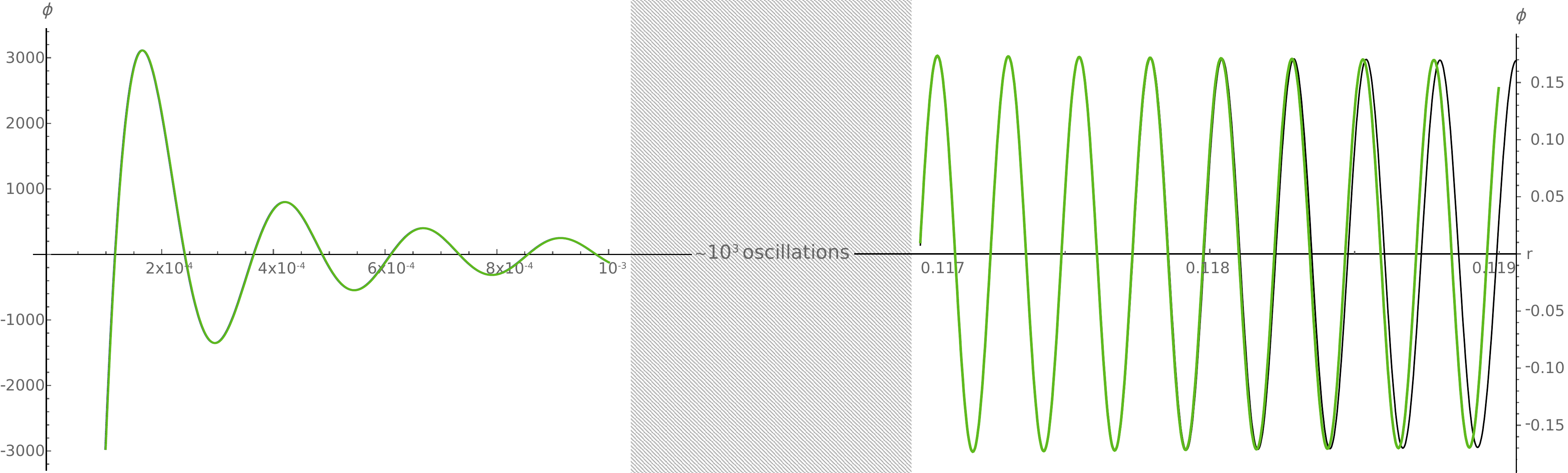}
\caption{\label{fig:scalarosc1} Full non linear evolution (color) and approximate solution (black) for the scalar field profile $\phi$ at the beginning (right) and end (left) of the last relaxation epoch. 
The fit reproduces correctly the order thousand oscillations that lye in the middle. The parameters correspond to Fig. \ref{fig:HairyBlackHoleProfile} (left).
}
\end{center}
\end{figure}

During the relaxation epoch corresponding to the last revival, we need to obtain an expression for the scalar field that do not relies on the validity of the WKB approximation. Such expression would allow us to understand the behavior of the solution as we leave the WKB epoch.  To obtain it, we notice that the gauge field is constant $h = h_n$ while the lapse function $N$ approaches zero  and the radial function $f$ grows linearly, but notably the product $Nf$ remains constant $Nf= b_n$. Plugging this information into the equations of motion for the scalar field \eqref{eq:scalar} we can solve it obtaining
\begin{equation}
\phi= \frac{c}r\, J_1\left(\frac{q(h_h-h_{n})}{b_n}\,r\right)+\frac{d}r\, Y_1\left(\frac{q(h_h-h_{n})}{b_n}\,r\right)\,.
\label{eq:scalarosc}
\end{equation}
where $c$ and $d$ are integration constants. By fitting them, we can approximate correctly the profiles for the scalar field over a long range of $r$. In particular, in Fig. \ref{fig:scalarosc1} we show the extremes of the regions where the oscillations begin and end, but the fit reproduces correctly the order thousand of oscillations that lye in the middle.


\subsection{Gauge field oscillations}
\label{subsec:gfo}

As the systems emerges from the WKB epoch into the last stage of its evolution, the solution \eqref{eq:scalarosc} diverges for small $r$ as $\phi=\phi_s/r^2$ with $\phi_s={2db_n}/{q(h_h-h_{n})\pi}$. This  correctly captures the behavior for the scalar field as we get into the small $r$ region.  In order to solve the remaining equation, notice that if we assume that $f=f_s/r^2$ and $rh'$ is bounded, then equation \eqref{eq:radial} implies that $fN$ is constant and then $N = N_s r^2$. This is consistent with equation \eqref{eq:lapse} as long as $h$ is bounded and $f_s=-\phi_s^2/3\alpha$. Then the only remaining equation is the one for the gauge field \eqref{eq:maxwell}, which reads
\bea
&&
r\left(r h'\right)'
=-6\alpha   q^2  (h-h_h)\,,
\label{eq:maxwellGFO}
\eea
which is integrated immediately as
\bea
h=h_h+h_s\cos\left(\sqrt{6\alpha}q\log (r/r_s)\right)
\eea
being $r_s$ and $h_s$ constants of integration.

\begin{figure}[htb]
\begin{center}
\includegraphics[width=8.1cm]{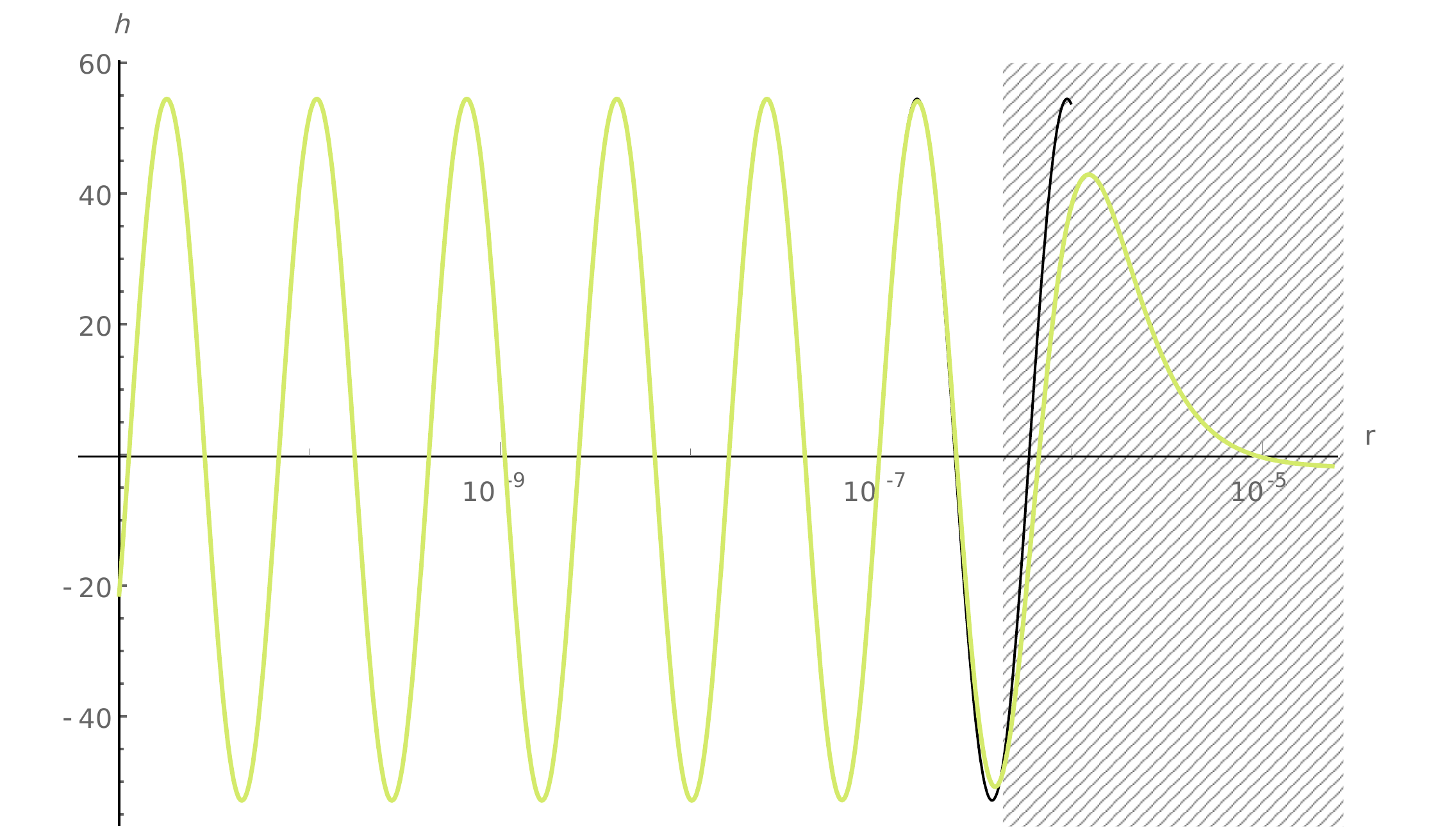}\hfill 
\includegraphics[width=8.1cm]{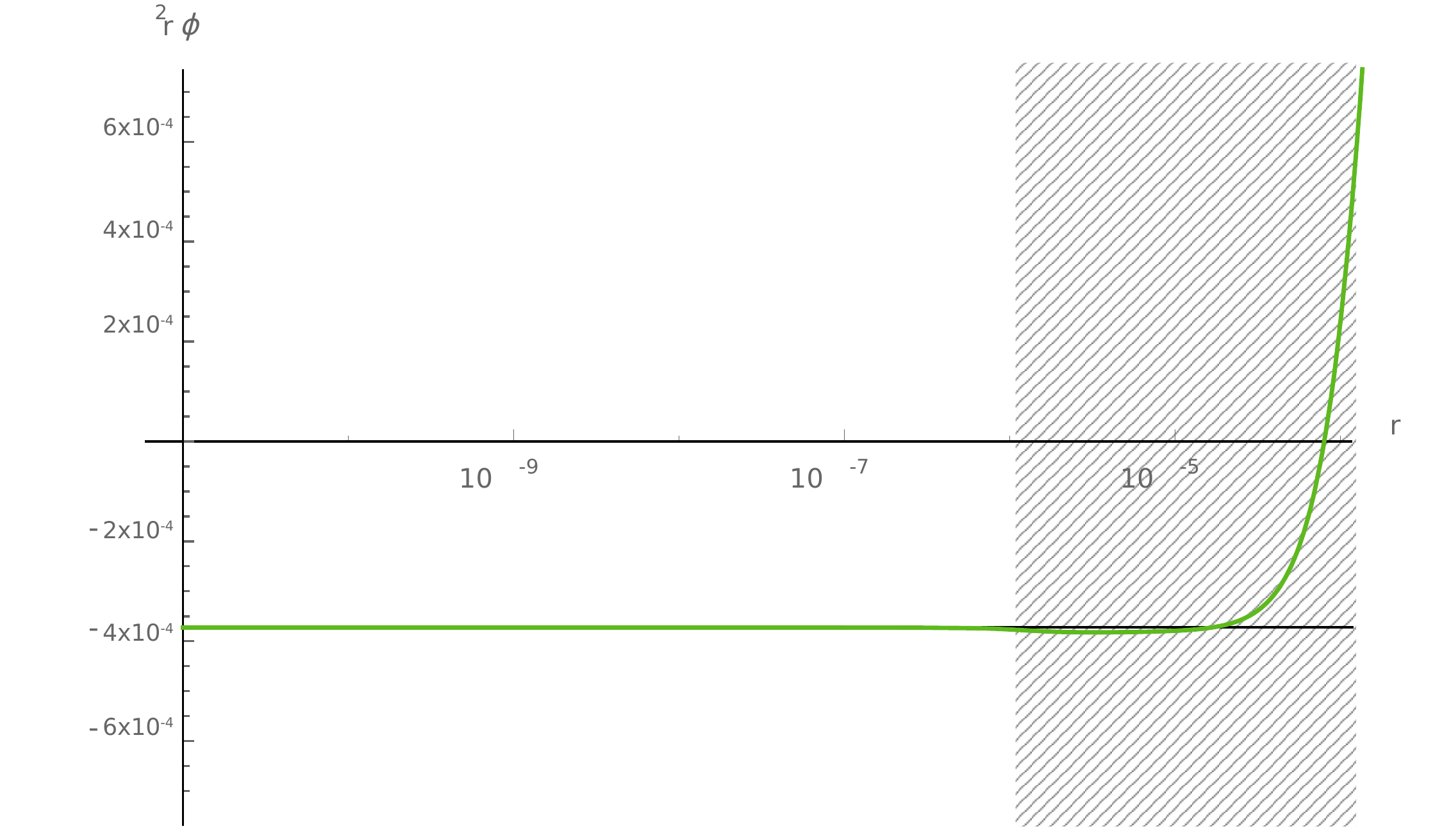}
\caption{\label{fig:gaugeosc1} Radial profiles for $h$ (left) and $r^2 \phi$ (right) on the gauge oscillations epoch. The colored curves correspond to the full non linear evolution while the black ones to the small $r$ approximation. The values of the parameters correspond to the black hole shown in Fig. \ref{fig:HairyBlackHoleProfile} (left).}
\end{center}
\end{figure}

{Notice that the resulting geometry is still singular, with a scalar curvature that diverges as $r^{-4}$ as we approach the center. To make further contact with previous literature \cite{Damour:2002tc,Hartnoll:2020rwq}, we change coordinates to $\tau={r^2}/{2\sqrt{-f_s}}$, so that the geometry near the singularity reads
\begin{equation}
ds^2\approx -d\tau^2+2\tau\,\sqrt{-f_s}\left(- f_s N_s^2 dt^2+ d\Omega_3^2\right)\,.
\end{equation}
Then the behavior near the singularity mimics a Kasner metric, with modified exponents due to the presence of the scalar field and the Gauss-Bonnet coupling \cite{Kirnos:2009wm}.}

\section{Discussion and Future directions}
\label{sec:end}

An interesting sequence of dynamical phase transitions appeared as we moved away from the horizon diving into the interior of the black hole. We managed to characterize these epochs by some salient features that can be easily read from the numerics and then reproduced with approximate solutions to the equations of motion. The analysis of the behavior of such phases suggest some possible future lines of research

\begin{itemize}

\item The relaxation phase appears as an intermediate regime for all the solutions we studied. Interestingly, for low enough temperatures, we observe revivals were the fields change abruptly to come again to this intermediate regime. This kind of behavior resembles a lot to the pre-thermal states in the Generalized Gibbs ensemble for quantum dynamic phases after a sudden quench \cite{moeck, mori,Cardy:2014rqa}. In our case, we might interpret the collapse of the Einstein-Rosen bridge as the quick playing the role of a quench.

\item After a series of revivals, the evolution ends into a final state that has similar features for all our solutions, characterized by the gauge fields oscillations. Comparing again with the quantum quench picture, one might think of it as the thermalized state. Once the system thermalizes, one would not expect any further dynamic phase transitions.

Notice that inside the black-hole the space-like coordinates are topologically $S_3\times R$. Hence one might argue that the revivals are associated to times where the dynamics are ruled by the $S_3$, while eventually the non-compact line and the non-linearities hit in and force the thermalization. We are not aware of such geometry driven phenomenology in the quantum quench literature, and it would be nice to realize it in a typical toy model, such as the Bose-Hubbard model.

\item Charged black holes often represent poor man's versions of Kerr black holes. Similar features happen in these two, but the charged black holes are much simpler to study since they are spherically symmetric. In this sense, it would be interesting to understand how would a scalar hair affect the interior of Kerr black holes \cite{Herdeiro:2014goa}.

\item By studying the spin 1 cousins of the present hairy black holes in the charged Proca theory \cite{Herdeiro:2016tmi,Brito:2015pxa,Garcia:2016ldc}, could shed light into how generic are the results presented here. These solutions have their asymptotically AdS counterparts \cite{Arias:2016nww}, which break rotation invariance for some values of the parameters \cite{Cai:2013aca}. This suggests that new interesting features may appear in such models.

\item Finally, references \cite{Hartnoll:2020fhc,Cai:2020wrp} present a simple and elegant proof for the non-existence of a second horizon in terms of an (almost) conserved charge that strongly constrains the dynamics between two horizons. Unfortunately as soon as we turn on the Gauss-Bonnet coupling $\alpha$ the proof gets ruined as no simple extensions are evident. { However, an interesting point is that, close to the collapse of the Einstein-Rosen bridge, the Gauss-Bonnet term decouples and the dynamics is completely controlled by Einstein gravity. This somewhat intriguing fact makes it natural to expect a similar phenomenology.} 

\end{itemize}
\section*{Acknowledgements}
\label{sec:acknowledgements}
We would like to thank Daniel Areán, Karl Landsteiner, Julio Oliva and Nico Nessi for correspondence. This work has been funded by the CONICET grants PIP-2017-1109 and PUE 084 ``B\'usqueda de Nueva F\'\i sica'', and UNLP grants PID-X791.

\end{document}